\documentclass[twocolumn]{aastex62}
\usepackage{natbib}
\shorttitle{Interpreting observations of TXS 0506+056}
\shortauthors{The MAGIC Collaboration}

\begin{document}

\title{The blazar TXS 0506+056 associated with a high-energy neutrino: insights into extragalactic jets and cosmic ray acceleration}

\author{S.~Ansoldi} \affil{Universit\`a di Udine, and INFN Trieste,
I-33100 Udine, Italy} \affil{Japanese MAGIC Consortium: ICRR, The University of Tokyo, 277-8582 Chiba, Japan; Department of Physics, Kyoto University, 606-8502 Kyoto, Japan; Tokai University, 259-1292 Kanagawa, Japan; RIKEN, 351-0198 Saitama, Japan}
\author{L.~A.~Antonelli} \affil{National Institute for Astrophysics (INAF), I-00136 Rome, Italy} 
\author{C.~Arcaro } \affil{Universit\`a di Padova and INFN, I-35131 Padova, Italy} 
\author{D.~Baack} \affil{Technische Universit\"at Dortmund, D-44221 Dortmund, Germany} 
\author{A.~Babi\'c} \affil{Croatian MAGIC Consortium: University of Rijeka, 51000 Rijeka, University of Split - FESB, 21000 Split,  University of Zagreb - FER, 10000 Zagreb, University of Osijek, 31000 Osijek and Rudjer Boskovic Institute, 10000 Zagreb, Croatia} 
\author{B.~Banerjee} \affil{Saha Institute of Nuclear Physics, HBNI, 1/AF Bidhannagar, Salt Lake, Sector-1, Kolkata 700064, India} 
\author{P.~Bangale} \affil{Max-Planck-Institut f\"ur Physik, D-80805 M\"unchen, Germany}
\author{U.~Barres de Almeida}
\affil{Max-Planck-Institut f\"ur Physik, D-80805 M\"unchen, Germany}
\affil{now at Centro Brasileiro de Pesquisas F\'isicas (CBPF), 22290-180 URCA, Rio de Janeiro (RJ), Brasil}
\author{J.~A.~Barrio} \affil{Unidad de Part\'iculas y Cosmolog\'ia (UPARCOS), Universidad Complutense, E-28040 Madrid, Spain} 
\author{J.~Becerra Gonz\'alez} \affil{Inst. de Astrof\'isica de Canarias, E-38200 La Laguna, and Universidad de La Laguna, Dpto. Astrof\'isica, E-38206 La Laguna, Tenerife, Spain}
% ...




\begin{abstract}
A neutrino with energy of $\sim$290 TeV, IceCube-170922A, was detected in coincidence with the BL Lac object TXS~0506+056 during enhanced gamma-ray activity, with chance coincidence being rejected at $\sim 3\sigma$ level.
We monitored the object in the very-high-energy (VHE) band with the MAGIC telescopes for $\sim$41 hours from 1.3 to 40.4 days after the neutrino detection. 
Day-timescale variability is clearly resolved. We interpret the quasi-simultaneous neutrino and broadband electromagnetic observations with a novel one-zone lepto-hadronic model, based on interactions of electrons and protons co-accelerated in the jet with external photons originating from a slow-moving plasma sheath surrounding the faster jet spine. We can reproduce the multiwavelength spectra of TXS 0506+056 with neutrino rate and energy compatible with IceCube-170922A, and with plausible values for the jet power of $\sim 10^{45} - 4 \times 10^{46} {\rm erg \ s^{-1}}$.
The steep spectrum observed by MAGIC is concordant with internal $\gamma\gamma$ absorption above $\sim$100 GeV entailed by photohadronic production of a $\sim$290 TeV neutrino, corroborating a genuine connection between the multi-messenger signals. In contrast to previous predictions of predominantly hadronic emission from neutrino sources, the gamma-rays can be mostly ascribed to inverse Compton up-scattering of external photons by accelerated electrons. The X-ray and VHE bands provide crucial constraints on the emission from both accelerated electrons and protons. We infer that the maximum energy of protons in the jet co-moving frame can be in the range $\sim 10^{14}$ to $10^{18}$ eV.
\end{abstract}



\section{Introduction} 
\label{sec:intro}
The birthplace of ultra-high-energy cosmic rays (UHECRs), the most energetic particles known in the Universe with energies exceeding $10^{18}$ eV, is a long-standing mystery \citep{Dawson:2017rsp}.
The observed distribution of their arrival directions in the sky favor a predominantly extragalactic origin \citep{Aab:2017tyv}.
Among the numerous candidate sources that have been proposed \citep{Hillas:1985is}, one of the most promising are active galactic nuclei (AGN) that eject powerful jets of magnetized plasma at relativistic velocities \citep{Biermann:1987ep}.
Blazars are AGN with their jets oriented close to the line of sight of the observer.  
Their predominantly non-thermal spectral energy distribution (SED) typically consists of two well-defined, broadly-peaked components. The one at lower energy, peaking in the infrared to X-ray bands, is due to synchrotron radiation by high-energy electrons. The other one at higher energy, peaking in the GeV-TeV gamma-ray range, is generally attributed to inverse Compton (IC) upscattering of low-energy photons by high-energy electrons.
Blazars can be categorized into two main subclasses. Flat spectrum radio quasars (FSRQs) are relatively luminous and display strong, optical-UV emission lines in addition to the non-thermal continuum. The low-energy seed photons for IC emission are likely dominated by thermal photons originating outside the jet, leading to external-Compton (EC) emission \citep{Madejski:2016oqg}. BL Lac objects are relatively less luminous and show only weak emission lines.
The low-energy seed photons for IC emission are often thought to be synchrotron photons within their jets, leading to synchrotron-self-Compton (SSC) emission.

The mechanism that accelerates electrons is also expected to accelerate protons and nuclei, potentially up to the energy range of UHECRs \citep{Bykov:2012ca}. These hadrons can interact with ambient matter or low-energy photons to generate gamma-rays, as well as neutrinos, as envisaged in hadronic and lepto-hadronic scenarios of blazar emission (for reviews see \citealt{Halzen:2016gng, Meszaros:2017fcs}). 
Neutrinos can be considered the ``smoking gun'' of hadron acceleration. In contrast to cosmic rays that may be deflected by intervening magnetic fields while propagating to the observer, photons and neutrinos are expected to point back to their origin in the sky, providing critical insights into the sources of cosmic rays.

In September 2017, the IceCube neutrino observatory revealed an event designated IceCube-170922A with 56.5\% probability of being a truly astrophysical neutrino. The best-fit reconstructed direction was at 0.1$^{\circ}$ from the sky position of the BL Lac object TXS~0506+056\footnote{The IceCube Collaboration, GRB Coordinates Network, Circular Service, No. 21916 (2017).}. The most probable energy was found to be 290 TeV (311 TeV), with the 90\% C.L. lower and upper limits being 183 TeV (200 TeV) and 4.3 PeV (7.5 PeV) respectively, assuming a spectral index of -2.13 (−2.0) for the diffuse astrophysical muon neutrino spectrum \citep{Aartsen:2014aef}. 
No additional excess of neutrinos with lower energy was found from the direction of TXS~0506+056 near the time of the alert \citep{eaat1378}.

Extensive follow-up observations revealed TXS~0506+056 to be active in all electromagnetic (EM) bands, most notably in GeV gamma-rays monitored by the {\it Fermi}-LAT (Large Area Telescope; \citealt{ATel:Fermi170922}), and in very-high-energy (VHE) gamma-rays above 100 GeV, detected for the first time from this object by the MAGIC (Major Atmospheric Gamma-ray Imaging Cherenkov) telescopes \citep{ATel:MAGIC170922}. The redshift has been recently measured to be $z=0.3365 \pm 0.0010$ \citep{Paiano:2018qeq}. 

The probability for a high-energy neutrino to be detected by chance coincidence with a flaring blazar from {\it Fermi}-LAT catalogs 
was found to be disfavored at a 3$\sigma$ confidence level, mostly due to the precise determination of the direction of IceCube-170922A \citep{eaat1378}.

These measurements offer a unique opportunity to explore the interplay between energetic photons, neutrinos and cosmic rays.\footnote{A potential association between a high-energy neutrino and an active blazar had been found in a previous study, but with much larger uncertainties in the positional and temporal coincidence \citep{Kadler:2016ygj}.}
Hereafter we interpret within a coherent scenario the available multi-messenger data presented in \cite{eaat1378}, together with additional MAGIC data for TXS 0506+056, under the assumption that the association of IceCube-170922A and the blazar in an active state is genuine, and that the neutrino and EM emission arise from the same region in the object. Given the appreciable probability of IceCube-170922A being of atmospheric origin, and that chance coincidence is rejected only at $\sim 3\sigma$ level, we note that the validity of the association is still an open question.

SED data strictly simultaneous with the neutrino event is not available for all wavelengths, and the duration of the active state responsible for the neutrino emission is uncertain. On the other hand, the data from Fermi reveals enhanced gamma-ray activity of TXS~0506+056 lasting several months. We consider physical models of the source that are compatible with the detection of one neutrino by IceCube during 0.5 years in the energy range reported in \cite{eaat1378}.
By invoking a dense field of low-energy photons originating outside the jet as targets for photo-hadronic interactions, the measured neutrino event can be interpreted consistently with the EM observations. 
Furthermore, while the detection of the $\sim 290$ TeV neutrino alone indicates acceleration of protons in the jet of this object to energies of at least several times $10^{15}$ eV, the combination with the EM data allow us to probe the maximum energy that they can attain.

\section{VHE gamma-ray and broadband emission of TXS 0506+056}
The rapid dissemination of the sky position of IceCube-170922A triggered an extensive multi-wavelength (MWL) campaign by many telescopes on ground and in space, from radio frequencies up to VHE gamma-rays. In the VHE band, TXS 0506+056 was followed up by several Imaging Atmospheric Cherenkov Telescopes (IACTs), with the earliest one starting a few hours after the neutrino alert \citep{eaat1378}. 

\subsection{MAGIC VHE gamma-ray Observations and Results}
MAGIC \citep{Aleksic:2014lkm} is a system of two 17m ground-based IACTs, operating in stereoscopic mode since fall 2009 at the Roque de los Muchachos Observatory 
($28.8^{\circ}$ N, $17.8^{\circ}$ W, 2200 m a.s.l.), on the island of La Palma, Canary Islands
 (Spain). 
The telescopes record Cherenkov light from extended air shower events starting from 30 GeV up to $\sim$100 TeV within a field of view of $\sim$10 square degrees.

\begin{figure}[]
\centering
\includegraphics[width=0.5\textwidth]{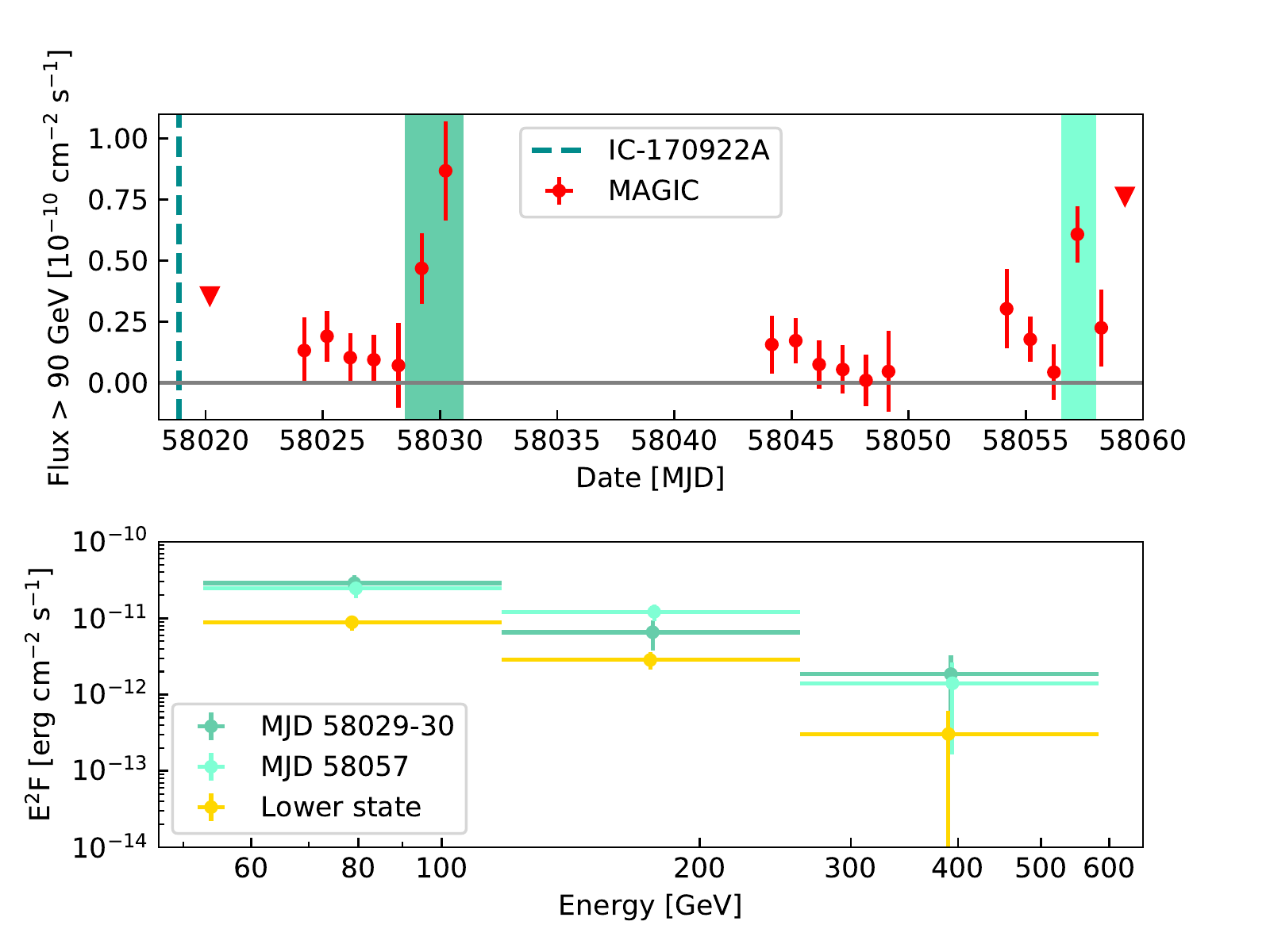}
\caption{{\bf Top:} Very high energy (VHE, E$>$90 GeV) gamma-ray light curve of the blazar TXS 0506+056 as measured by MAGIC. Data from MDJ 58020 to MJD 58030 has been already presented in \citep{eaat1378}. The colored boxes mark the two periods of enhanced emission during MJD 58029-30 and MJD 50857. The triangles at MJD 58020 and 58059 are 2 $\sigma$ upper limits. The dashed blue line indicates the arrival time of the high energy neutrino event IC-170922A (MJD 50818).
{\bf Bottom:} Spectral energy distribution of the blazar TXS 0506+056 as measured by MAGIC in different observation periods. 
}
 \label{fig_LC}
\end{figure}

MAGIC observations of the sky position of IceCube-170922A from September 24 (MJD 58020) till October 4, 2017 (MJD 58030) yielded the first  detection of a VHE signal from an astrophysical object in a direction compatible with that of the high-energy neutrino, as described in \cite{eaat1378}. In this work, we present additional MAGIC data on TXS 0506+056 collected until November 2, 2017 (MJD 58059), adding up to a total exposure of $\sim$47 hours. After data quality cuts based on the atmospheric conditions, telescope response and stereo event rate, about 41 hours of data were selected for further analysis, employing the standard MAGIC analysis framework \texttt{MARS}. 
Parameter cuts optimized for low energies and tuned with data from the Crab Nebula were used \citep{Aleksic:2014lkm}. Results for gamma-rays above 90 GeV are given in Figure \ref{fig_LC} and Table \ref{tab_obs}. 

The VHE gamma-ray flux is variable, increasing by a factor of up to $\sim$6 within one day.
The probability of a constant flux is less than 0.3\%.
Two periods of enhanced VHE gamma-ray emission can be clearly distinguished: one on MJD 58029-58030 (Oct 3-4, 2017, already reported in \cite{eaat1378}) and a second one on MJD 58057 (Oct 31, 2017).
Detection of TXS~0506+056 with MAGIC was also achieved (at the level of 5.7$\sigma$) by stacking all data with a similarly low VHE gamma-ray flux level (F(E$>$90 GeV) $<$ 4.5 $\times$ 10$^{-11}$ cm$^{-2}$ s$^{-1}$) measured outside these two periods. This allowed us to distinguish between the states with high and low VHE gamma-ray emission. Note that, the low state is unlikely to exemplify the typical quiescent state of the source according to the historical data in Figure \ref{fig_SED} c).

We could not reveal any difference in the spectral shape in periods with different flux levels or data samples. The measured differential photon spectrum can be described over the energy range of 80 GeV to 400 GeV by a simple power law $dN/dE \approx E^{\gamma}$, with a spectral index ranging from (-4.0 $\pm$ 0.3) to (-3.5 $\pm$ 0.4), see Table \ref{tab_spec}. 
This is significantly steeper than the spectrum measured by Fermi-LAT quasi-simultaneously, with spectral index (-2.0 $\pm$ 0.2) in the energy range 100 MeV to 300 GeV.
The systematic uncertainties of MAGIC spectral measurements can be divided into: $<$ 15\% in energy scale, 11-18\% in flux normalization and $\pm$ 0.15 for the energy spectrum power-law slope \citep{Aleksic:2014lkm}. In a view of the lack of evidence for spectral variability within the data sets reported here, we adopt the spectral index derived from the stacked data as a common value for the lower VHE state.

\subsection{Multi-wavelength SED}
Comprehensive MWL coverage of TXS~0506+056 during MJD 58019-30 is reported in \cite{eaat1378}. For this work, we analyzed additional open access MWL data with public tools during the period from MJD 58019 to MJD 58059 from the following instruments: KVA \citep{2016AA...593A..98L},
UVOT and XRT onboard the Neil Gehrels Swift observatory (\textit{Swift}) \citep{2005SSRv..120...95R}, NuSTAR \citep{2013ApJ...770..103H} and {\it Fermi}-LAT \citep{2009ApJ...697.1071A}.
For each waveband considered, data within 24 hours around the MAGIC observations were combined and considered as quasi-simultaneous for the modeling.
Figure \ref{fig_SED} a) shows the SED restricted to the first period of VHE gamma-ray enhanced emission observed from MJD 58029 to 58030. Good coverage of simultaneous MWL data is also found for the MAGIC observations at the lowest flux in Figure \ref{fig_SED} b). The second period of enhanced VHE gamma-ray flux (MJD 58057) has limited MWL coverage and is not used for the modeling below. 

Comparison of the light curve in Figure \ref{fig_LC} with that measured by {\it Fermi}-LAT \citep{eaat1378} suggests the low VHE state to be more representative of the VHE average emission during the six months of enhanced GeV flux. 
On the other hand, the VHE emission of the flare state may be considered an upper bound.
By compiling broadband SEDs quasi-simultaneous with the MAGIC observations, we extrapolate such considerations to all other wavelengths. 
The obtained SEDs appear typical of BL Lac objects with similar luminosities \citep{Tavecchio:2009zb}.

\section{Interpreting the broadband and neutrino emission}
Production of high-energy neutrinos in astrophysical environments is expected to occur mainly through the decay of charged pions produced in inelastic collisions between high-energy protons and ambient target particles \citep{Halzen:2016gng,Meszaros:2017fcs}, which can be either matter ($pp$ interactions) or low-energy photons ($p\gamma$ interactions). In AGN jets that typically consist of low-density plasma,
the latter is generally the favored channel \citep[e.g.][]{Mannheim:1995mm}.
Photo-hadronic production of neutrinos with energy $E_\nu \sim 290$ TeV requires interactions between protons with energy $E_p \simeq 6$ PeV and target photons with energy exceeding the corresponding threshold, $\epsilon \simeq m_\pi m_p c^4 / E_p \simeq$ 440 eV, that is, in the UV to soft X-ray band.

FSRQs are generally considered to be promising sources of  high-energy neutrinos, on account of the inferred high density of target photons for the $p\gamma$ channel \citep{Atoyan:2001ey, Murase:2014foa}.
In contrast, BL Lac objects such as TXS~0506+056 have generally been thought to be inefficient neutrino emitters \citep{Murase:2014foa}, due to the relatively low density of UV to soft X-ray synchrotron photons expected inside the jet. 
Thus, the association of a high-energy neutrino with TXS 0506+056 is not trivial to interpret.
Models in which hadronic emission components are more prominent for the gamma-rays may allow more neutrino emission, at the expense of relatively large values for the power in accelerated protons that may not be so favorable from an energetics perspective (see e.g. Cerruti et al., in preparation, for the specific case of TXS 0506+056).

Higher efficiency of neutrino production without invoking high values of the proton power may still be achievable by considering target photon fields external to the jet. 
One such scenario 
in which external photons naturally occur 
for BL Lac objects is the so called spine-layer or jet-sheath scenario, as suggested
by \citet{Tavecchio:2014eia} (see also \citealt{Righi:2016kio}).
This model postulates structured jets, consisting of a faster core (or spine) surrounded by a slower sheath (or layer). Such a velocity structure is supported by observational and phenomenological evidence, as well as numerical simulations \citep{Tavecchio:2014eia}.
Due to significant
relative motion between the two structures, the density
of photons originating from the slower sheath appears highly
boosted in the frame comoving with the faster jet \citep{Ghisellini:2004ec}, exceeding
that of the synchrotron photons arising within the jet,
and thus enhancing the rate of $p\gamma$ reactions and consequent
neutrino emission. These external photons serve not only as targets for $p\gamma$ interactions, but also as seed photons for IC upscattering, so that EC emission may also play an important role.

As long as neutrinos are produced via the $p\gamma$ channel, an important consequence is the associated absorption of gamma-rays via $\gamma\gamma$ interactions \citep{WaxmannBahcall:1999,Dermer:2007co}. Since the cross section for $\gamma\gamma$ interactions is on average $\sim 10^3$ times larger than for $p\gamma$ interactions, the presence of target photons that provide a reasonable production efficiency of neutrinos at a given energy inevitably implies strong gamma-ray absorption above a corresponding threshold.   
Detection of such a spectral feature constitutes a critical test of photohadronic neutrino production as well as the neutrino-blazar association itself.

\subsection{Model description}
\label{sec:model}
Hereafter we model the broadband EM and neutrino emission from TXS~0506+056, building on the jet-sheath scenario of \citet{Ghisellini:2004ec,Tavecchio:2014eia}. These earlier studies focused on the leptonic emission and/or the neutrino emission and lacked explicit treatment of hadronic radiative processes. Here we present the first extension of the jet-sheath model to account for all relevant hadronic processes, as well as a first comparison of the model with multi-messenger observations of a particular object.

The radiative output is calculated in the reference frame comoving with the jet flow, assuming isotropic distributions for both electrons and protons. This is then transformed to the observer frame via the Doppler factor $\delta$, which accounts for effects due to the jet's relativistic motion 
\citep{Madejski:2016oqg}.

The calculation of the leptonic emission processes, which include synchrotron, SSC and EC emission, follows \cite{Ghisellini:2004ec}. We note that in EC models such as the jet-sheath scenario, the angular distribution of IC seed photons in the jet comoving frame will be highly anisotropic due to relativistic boosting. This leads to a more narrowly peaked beaming pattern for the EC emission compared to the synchrotron and SSC emission
\citep{1995ApJ...446L..63D, Ghisellini:2004ec}.

We account for all relevant hadronic radiative processes, including photo-meson-induced cascade emission \citep{Mannheim:1993jg}, Bethe-Heitler (BH) pair cascade emission \citep{Petropoulou:2014rla}, and synchrotron radiation from protons \citep{Aharonian:2000pv} and muons. The former is the most important component in our model. For protons of sufficiently high energy, the gamma-rays resulting from photo-meson reactions can be energetic enough to undergo electron-positron pair production interactions with low-energy photons ($\gamma\gamma\to e^{+}e^{-}$), thereby triggering secondary pair cascades \citep{Mannheim:1993jg}. This process redistributes the energy of photons down to lower energies until it falls below the threshold for $\gamma\gamma$ interactions and escape,
generally resulting in a spectrum with roughly equal power over a broad energy range (from optical to VHE for the cases shown below).
From the spectral properties of the external radiation fields considered here, $\gamma\gamma$ transparency is expected below a few hundreds of GeV, the energy band best accessible to MAGIC. 
The direct products of $p\gamma$ reactions including neutrinos and photons are described using the analytical treatment of \citet{Kelner:2008ke}. The secondary pair cascading processes were implemented following the formalism of \cite{Boettcher:2013wxa} and extending it to include IC in addition to synchrotron processes for the pairs.
An additional hadronic component arises from BH pair production ($p\gamma\to p e^{+}e^{-}$), which can be observationally relevant in some cases.
Gamma-ray emission via synchrotron radiation of protons or muons is mostly unimportant for the range of magnetic fields and proton energies considered here, except for a limited region of parameter space. 
The results presented here have been cross-checked with the codes developed in \cite{Cerruti:2014iwa} and \cite{Zech:2017lma}.

As mentioned above for EC emission, in the jet comoving frame, target photons from the sheath should appear highly anisotropic. So far, there has been little discussion in the literature concerning photohadronic neutrino and cascade emission in anisotropic photon fields. The anisotropy may induce an effect analogous to EC for the neutrino emission, but the non-trivial propagation of charged pions and muons in magnetic fields before their decay complicates the picture. On the other hand, the accompanying cascade emission is expected to be mostly isotropic, since the secondary $e^{\pm}$ pairs are likely isotropized by magnetic fields as they degrade in energy. Thus, the neutrino emission may be more narrowly beamed compared to the cascade emission. For the viewing angles and magnetic fields considered here, we estimate that this effect may enhance the observed neutrino flux by a factor of $\sim$2-3 relative to the case assuming isotropy. However, this is not explicitly included in our calculations, as a full investigation of the effect requires Monte Carlo simulations that are beyond the scope of this paper.

The main parameters of the model specifying the jet environment (magnetic field, Doppler factor, size of acceleration/emission region) and the particle populations (density and energy distribution of electrons and protons) are adjusted to consistently reproduce the MWL data and the observed neutrino event rate. 
The absorption of gamma-rays during their propagation to the observer caused by $\gamma\gamma$ interactions with the extragalactic background light is modeled following \cite{Dominguez:2011a}, and is expected to be minor at the measured redshift of TXS 0506+056 at energies below 400 GeV ($\sim$10\% at 400 GeV and $\sim$50\% at 4 TeV).

\subsection{Results}
\begin{figure*}[]
\centering
\includegraphics[width=1.0\textwidth]{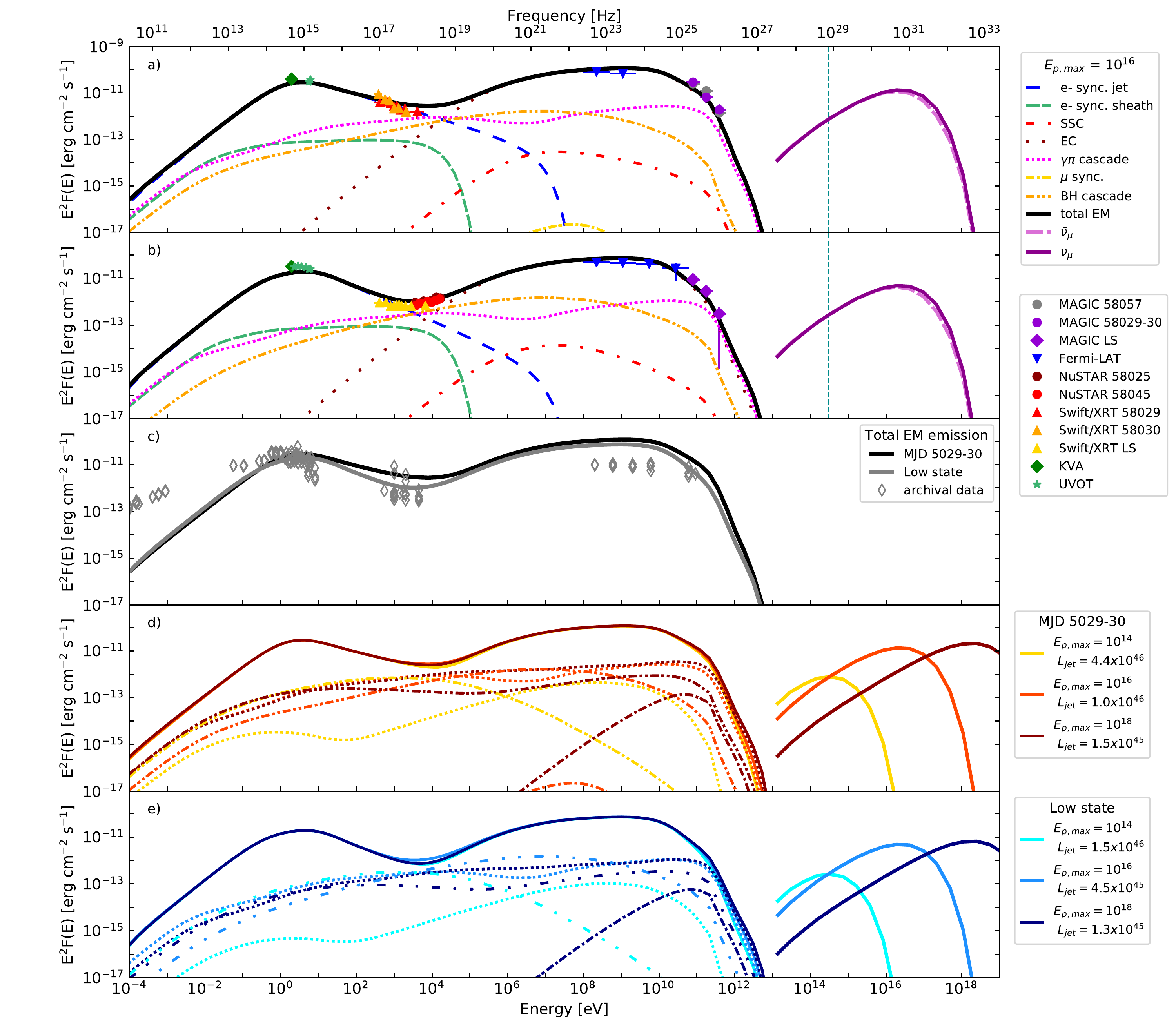} 
\caption{Spectral energy distribution for the enhanced VHE gamma-ray emission state ({\bf a)}, MJD 58029 to 58030) and the lower VHE gamma-ray emission state (LS, {\bf b)}) modeled with the jet-sheath scenario with $E_{\rm p,max}=10^{16}$ eV. Symbols corresponding to data-points from different facilities and observation epochs are described in the legend.
The curves represent individual emission components while the thick black curve shows the total predicted emission. The leptonic emission from the jet includes synchrotron (blue loose-dashed), synchrotron-self-Compton (SSC, red loose-dash-dotted), and external Compton (EC) emission (dark red loose-dotted). Synchrotron emission from the sheath is denoted by the green dense-dashed line.
The hadronic emission components are photo-meson-induced cascade (purple dense-dotted), Bethe-Heitler pair cascade (dark yellow double-dot-dashed) and muon-synchrotron (yellow dash-dotted).
Predicted (anti-)neutrino spectra are marked by (light-)magenta (dashed) solid lines, the blue vertical line shows the energy $\sim$290 TeV of the observed neutrino. A comparison of the two solutions is also shown with the archival data from ASDC ({\bf c)}).
Results for different values of $E_{\rm p,max}$ are compared for the 
enhanced VHE gamma-ray emission state ({\bf d)}, MJD 58029 to 58030) and the lower VHE gamma-ray emission state (Low state, {\bf e)}).}
\label{fig_SED}
\end{figure*}

Figure \ref{fig_SED} shows results for models corresponding to the low and high states (see details in the caption). Table \ref{tab_param} lists their parameters (all in the jet comoving frame): jet magnetic field $B$, parameters specifying the electron distribution (a broken power-law with minimum energy $E_{\rm min}$, break energy $E_{\rm br}$ and maximum energy $E_{\rm max}$, and spectral indices $n_1 = 2$ and $n_2$), 
and the energy densities in relativistic electrons $U_e$,
magnetic fields $U_B$,
and relativistic protons $U_p$.
The jet power is evaluated as $P_{\rm jet} = \pi R^2 \Gamma^2 c (U_e + U_p + U_B)$, for which the individual contributions from relativistic electrons $P_e$, magnetic fields $P_B$, and relativistic protons $P_p$ are also given in Table \ref{tab_param} \citep{Ghisellini:2009fj}.
For the two states, we assume the same values for the jet bulk Lorentz factor $\Gamma_j = 22$, viewing angle $\theta_{view} = 0.8 \degr$ (implying $\delta \simeq 40$), sheath bulk Lorentz factor $\Gamma_s = 2.2$, and the radius of the emission region $R = 10^{16}$ cm (corresponding to a variability timescale of one day). Note that we take $\theta_{view} < 1/\Gamma_j$, where the EC emission appears more luminous relative to the synchrotron, SSC and hadronic cascade emission than at $\theta_{view} \sim 1/\Gamma_j$, the viewing angle typically assumed in the literature, by virtue of the former's more narrowly peaked beaming pattern \citep{1995ApJ...446L..63D}. The energy distribution of relativistic protons is assumed to be a simple power-law with spectral index 2 and an exponential cutoff at maximum energy $E_{\rm p,max}$. We limit $B$ to a range such that muons and pions produced in hadronic interactions do not suffer significant energy losses before decaying.
We explore a range of $E_{\rm p,max}$ suitable for reproducing the observations. An upper limit of $E_{\rm p,max} \simeq 10^{18}$ eV is expected as the theoretical maximum allowed by plausible mechanisms of particle acceleration (see Section \ref{UHECR}).  
A lower limit of $E_{\rm p,max} \simeq 10^{14}$ eV in the comoving frame is imposed to allow production of a neutrino with observed energy $\sim$290 TeV, since lower values of $E_{\rm p,max}$ imply significantly less $p\gamma$ interactions above the corresponding energy threshold. Compensating this with higher proton power is not feasible, as the latter is limited by X-ray constraints on BH cascade emission, as well as jet energetics arguments.

Despite the large number of model parameters, the combined constraints from the EM and neutrino channels are quite powerful in assessing acceptable regions of the parameter space. Particularly strong constraints on the broadband hadronic cascade emission, and therefore on $E_{\rm p,max}$ in conjunction with the primary proton luminosity, come from
the combination of the X-ray and gamma-ray bands, which also constrain the spectrum of primary leptons.
The low state SED in Figure \ref{fig_SED} b) is more tightly constraining, thanks to the additional data from the {\it NuSTAR} satellite, which cover the range in between the two main SED components.

VHE gamma-ray data from MAGIC provide several important constraints for the modeling. First, the size of the emission region is limited by the observed day-timescale variability, which is unresolvable in the {\it Fermi}-LAT data. Second, the VHE spectrum depends sensitively on the energy distribution of primary electrons, in particular its break and maximum energies. Finally and most crucially, strong spectral steepening due to internal $\gamma\gamma$ absorption is robustly predicted in the VHE band, as an inevitable byproduct of photohadronic production of a $\sim$290 TeV neutrino. The fraction of multi-PeV parent protons that undergo $p\gamma$ interactions to produce $\sim$290 TeV neutrinos is $\sim10^{-4}$.
This implies that for the target photons in such $p\gamma$ interactions, the $\gamma\gamma$ optical depth is $\sim 0.1$ at the corresponding threshold $E_{\gamma\gamma} \simeq m_e^2 c^4 /\epsilon \simeq (20 m_e^2 / m_\pi m_p) E_\nu \simeq$ 12 GeV ($E_\nu$ / 290 TeV).
As the density of target photons is roughly inversely proportional to $\epsilon$ (Fig. 2), significant $\gamma\gamma$ absorption is expected above $\sim$ 100 GeV. The steep spectrum measured by MAGIC relative to {\it Fermi}-LAT indeed matches this expectation, providing a unique confirmation of the one-zone lepto-hadronic interpretation, the $p\gamma$ production channel, as well as the association between the neutrino and the blazar.

A good fit to the data is found for an intermediate value of $E_{\rm p,max}=10^{16}$ eV, which also yields the highest predicted neutrino event rates. Both higher and lower values of $E_{\rm p,max}$, in conjunction with X-ray and gamma-ray constraints, still allow acceptable solutions, with lower neutrino rates, as shown in Figure \ref{fig_SED} d) and e).

The muon neutrino fluxes predicted in our model are shown in Figure \ref{fig_SED}, considering neutrino oscillations into equal fractions among the three flavors during their propagation \citep{Aartsen:2015flavor}.
Convolved with the neutrino effective area reported in \cite{eaat1378}, we predict 90\% confidence level lower and upper limits to the muon neutrino energy of 206 TeV and 6.3 PeV for $E_{\rm p,max}=10^{16}$ eV.
This predicted range is in good agreement 
with the observed one, namely 183 TeV (200 TeV) and 4.3 PeV (7.5 PeV) for a spectral index of -2.13 (2.0) \citep{eaat1378}.
The expected detection rate of muon neutrinos in the above energy interval for the models in Table \ref{tab_param}
are about 0.17 events in 0.5 years for the higher VHE emission state, and 0.06 events in 0.5 years for the lower VHE emission state. These numbers are conservative in that they do not account for a potential contribution to the event rates due to interactions of tau-neutrinos that induce muons with a branching ratio of 17.7\%. 
Both rates are in agreement with the detection of the single neutrino during the period of enhanced GeV gamma-ray emission \citep{eaat1378}. 
As mentioned in Section \ref{sec:model}, the neutrino fluxes may be higher by a factor of $\sim 2-3$ if the anisotropy of the target photon field for $p\gamma$ interactions is properly taken into account.
Note that, the neutrino flux is also constrained by limits on the hadronic emission from the EM observations, most effectively from the X-ray and VHE bands.

In agreement with many previous studies of the broadband emission of blazars, the SEDs here comprise mostly leptonic emission. In particular, the second SED component peaking in the GeV band is largely accounted for by EC emission, for which the enhanced luminosity relative to the other emission components at $\theta_{view} < 1/\Gamma$ is important for achieving acceptable fits.
 This is in contrast to some earlier studies of neutrino emission from blazars that predicted predominantly hadronic gamma-rays. Our models show that this is not necessarily the case, and neutrino emission can be commensurate with the widely accepted view of primarily leptonic gamma-ray emission from blazars.
This implies that the values of the parameters that suitably reproduce the SEDs (in particular magnetic field, Doppler factor, electron density) are in the range commonly derived for BL Lac objects similar to TXS~0506+056 based on purely leptonic models \citep{Ghisellini:2009fj}.
On the other hand, the photo-meson cascade and BH components must generally be subdominant. 

From the comparison of the model parameters for the low and high states, the enhanced emission can be attributed to changes in the energy distribution of the electrons in the jet (harder high-energy slope and larger average energy) and increased power of the proton population. Such variability is quite consistent with the behavior commonly displayed by BL Lac objects during active states.
In both cases, the power of the jet is in the range $10^{45}$ to $4\times 10^{46}$ erg/s. These values are somewhat larger than that derived for BL Lac objects through purely leptonic models that do not account for accelerated protons \citep{Ghisellini:2009rb}.
The ratio of energy density in protons to electrons is $\sim 3600$ and $\sim 1700$ for the high and low states respectively, similar to that inferred in various astrophysical environments.

From the combined EM and neutrino data, we infer that the maximum energy of protons in the blazar emission region
is in the range $10^{14}-10^{18}$ eV in the comoving frame.

\section{Implications for UHECR acceleration in AGN jets}
\label{UHECR}
As shown above, the multi-messenger observations of TXS~0506+056 can be consistently interpreted by considering the acceleration of electrons and protons in the same region of the jet, which interact with a conceivable source of underlying external photons. The detection of the single neutrino may provide the first compelling evidence for proton acceleration in AGN jets, while the MWL observations constrain key parameters such as the magnetic field and the Doppler factor of the relevant region. With minimal additional considerations from plausible theories of particle acceleration, we are in a unique position to infer the spectral distribution of protons in AGN jets, and discuss their viability as sources of UHECRs.

Various mechanisms have been proposed for accelerating UHECRs in AGN jets, involving strong shocks, magnetic reconnection, shear flows, etc. \citep{Bykov:2012ca,Sironi:2015oza}.
Independent of the specific
mechanism, as viewed in the comoving frame of the jet,
the timescale for accelerating particles up to energy E in a relativistic
flow can be expressed as $t_{\rm acc} = 10 \eta E / (Z e B c)$, where
$e$ is the elementary charge, $c$ is the velocity of light, and $\eta$
is a parameter that characterizes the properties of magnetic
disturbances responsible for the acceleration. 
For mildly relativistic 
shock velocities as expected in AGN jets, $\eta \simeq$ 1 corresponds
to the so-called Bohm limit whereby the turbulence is
fully developed and allows the fastest possible acceleration.
To be compared are timescales for loss processes due to adiabatic
expansion $t_{\rm ad} = 2 R/c$, synchrotron radiation $t_{\rm syn}(E)$,
photo-meson production $t_{\rm \gamma \pi}(E)$, and BH pair
production $t_{\rm \gamma e^{\pm}}(E)$ (e.g. \citealt{Cerruti:2014iwa}). The latter two depend on the spectra of
the target photon fields, and can be evaluated separately for
the internal photons (synchrotron from accelerated electrons)
and external photons considered here following \cite{Petropoulou:2014rla}. The maximum attainable proton energy $E_{\rm p, max}$ can be estimated by balancing $t_{\rm acc}(E)$ with $Z=1$ and the shortest loss timescale, $t_{\rm loss} = \min[t_{\rm ad}, t_{\rm syn}, t_{\rm \gamma \pi}, t_{\rm \gamma e^{\pm}}]$. Moreover the particle’s gyro-radius $r_g = E/ZeB$ should not exceed $R$ so that it remains confined in the acceleration region.

\begin{figure}[h]
\centering
\includegraphics[width=0.5\textwidth]{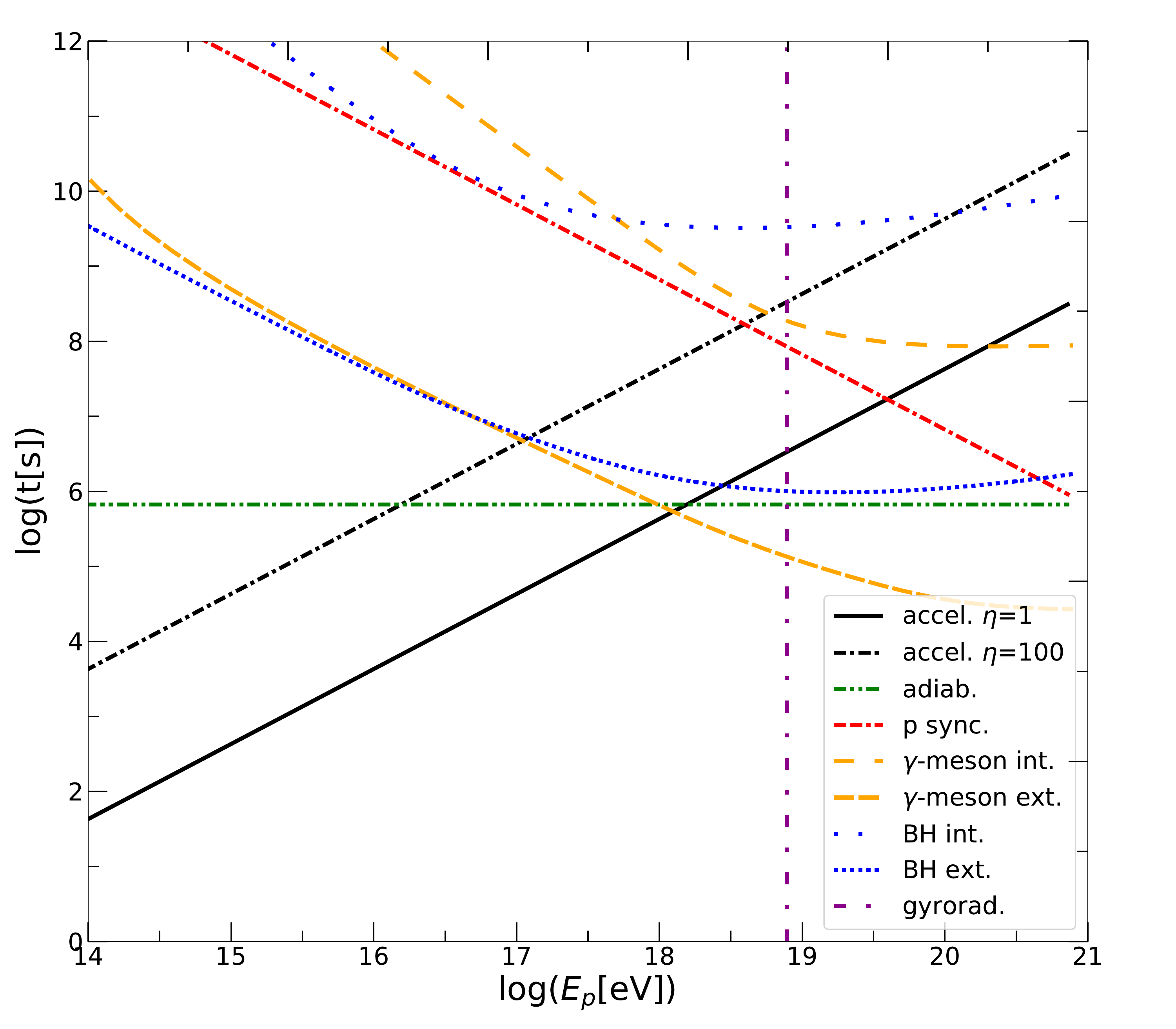}
\caption{Timescales for proton acceleration ($t_{\rm acc}$ with $Z = 1$) when $\eta = 1$ and $\eta = 100$, compared with those for adiabatic loss $t_{\rm ad}$, synchrotron loss $t_{\rm syn}$, photo-meson loss $t_{\rm \gamma \pi}$, and BH pair production loss $t_{\rm \gamma e^{\pm}}$, as functions of the proton energy $E_p$ in the comoving frame, for the high state. Also shown is the energy $E_{p,g} = eBR$ where the proton gyro-radius $r_g = R$.}
\label{fig_timescale}
\end{figure}

Figure \ref{fig_timescale} shows the comparison of the timescales 
and the gyro-radius constraint for the model of the high state discussed above. For  $\eta \simeq 1$, we see that protons can be accelerated up to $E_{\rm p,max} \simeq 10^{18}$ eV, limited by adiabatic and photo-meson losses on external photons. All values of $E_{\rm p,max} \la 10^{18}$ eV discussed above can be entirely compatible, with lower values of $E_{\rm p,max}$ corresponding to the balance of $t_{\rm ad}$ and $t_{\rm acc}$ with larger values of $\eta$ that may reflect lower levels of turbulence (e.g. $E_{\rm p,max} \simeq 10^{16}$ eV for $\eta =100$; see Fig.3). 

To contribute to the observed UHECRs, particles must escape from the acceleration region into ambient intergalactic space. 
The energy of escaping protons ${\cal E}_{\rm p,max}$ as seen by an observer will appear boosted by the relativistic bulk motion of the jet so that ${\cal E}_{\rm p,max} = \Gamma_j f_{\rm el} E_{\rm p,max}$ on average, where $f_{\rm el} \le 1$ is a factor that accounts for energy loss during the escape process. 
For our models with $\Gamma_j = 22$ and $E_{\rm p,max} \simeq 10^{14} - 10^{18}$ eV, ${\cal E}_{\rm p,max} \simeq 2 \times 10^{15} - 2 \times 10^{19}$ eV, as long as  $f_{\rm el} \sim 1$. The higher end of this range would be well within the domain of UHECRs. If nuclei heavier than protons are also accelerated in the same conditions, even higher energies may be possible.
 
\section{Discussion and Conclusions}
The combined MWL and neutrino observations of TXS~0506+056 can be interpreted consistently in terms of electrons and protons that are accelerated in the same region of the jet and interact with external low-energy photons, originating from the slower sheath surrounding the faster jet spine.
Such jet-sheath structure is independently well motivated. The enhanced luminosity for EC emission relative to the other emission components at small viewing angles is important for adequately reproducing the MWL SEDs. Most notably, the prominent spectral steepening observed at $\sim$100 GeV by MAGIC confirms the internal $\gamma\gamma$ absorption that is robustly expected as a consequence of $p\gamma$ production of a $\sim$290 TeV neutrino. Thus, our interpretation reinforces the association between the multi-messenger signals. While the bulk of the gamma-rays are inferred to be external Compton emission from electrons, a non-negligible contribution can arise from cascade emission induced by protons, most notably in the hard X-ray and VHE gamma-ray bands.

Jet structure as envisaged in the jet-sheath scenario can be conclusively tested with future very long baseline interferometry (VLBI) measurements with high angular resolution. We note that an alternative source of external photons in BL Lac objects may be provided by radiatively inefficient accretion flows around the central supermassive black holes (Righi et al, in preparation).

An interesting question concerns the potential contribution of the entire BL Lac population to the diffuse neutrino emission detected by IceCube \citep{Aartsen:2016lir}. In principle, BL Lac objects and FSRQs could account for a sizable fraction of the observed intensity \citep{Tavecchio:2014eia,Righi:2016kio}. The association of IceCube-170922A with the flaring phase of TXS~0506+056 is consistent with this scenario, although the relative rarity of these sources may imply additional constraints on their contribution \citep{Murase:2016gly}.

The inferred maximum energy of protons in the blazar emission region may be consistent with an important contribution to UHECRs
from protons and/or heavy nuclei accelerated in the blazar region. Although stronger constraints are not possible with a single observed neutrino event, future multi-messenger observations of blazars will offer a more critical probe of UHECR acceleration in the inner regions of AGN jets.

\begin{table}[]
\caption{MAGIC flux measurements. Enhanced emission states as discussed in the text are marked in bold.}
\label{tab_obs}
\centering
\begin{tabular}{lcc}
\hline
Date & Effective time & Flux $>$ 90 GeV  \\
MJD & [hours] & [10$^{-11}$ cm$^{-2}$ s$^{-1}$] \\
\hline
58020.18 & 1.1 & $<$ 3.56  \\ 
58024.21 & 1.2 & 1.3 $\pm$ 1.3  \\ 
58025.18 & 2.9 & 1.9 $\pm$ 1.0  \\ 
58026.17 & 3.0 & 1.0 $\pm$ 1.0   \\ 
58027.18 & 2.8 & 0.9 $\pm$ 1.0  \\ 
58028.23 & 0.8 & 0.7 $\pm$ 1.7  \\ 
{\bf 58029.22} & 1.3 & 4.7 $\pm$ 1.4  \\ 
{\bf 58030.24} & 0.6 & 8.7 $\pm$ 2.0  \\ 
58044.16 & 1.9 & 1.6 $\pm$ 1.2 \\
58045.18 & 3.2 & 1.7 $\pm$ 0.9 \\
58046.18 & 3.1 & 0.8 $\pm$ 1.0 \\
58047.19 & 2.7 & 0.6 $\pm$ 1.0 \\
58048.18 & 2.3 & 0.1 $\pm$ 1.0 \\
58049.14 & 1.0 & 0.5 $\pm$ 1.7 \\
58054.18 & 0.8 & 3.0 $\pm$ 1.6 \\
58055.19 & 2.9 & 1.8 $\pm$ 0.9 \\
58056.20 & 2.3 & 0.4 $\pm$ 1.1 \\
{\bf 58057.20} & 2.7 & 6.1 $\pm$ 1.2 \\
58058.22 & 1.6 & 2.3 $\pm$ 1.6 \\
58059.23 & 0.3 & $<$7.6 \\ 
\hline
 \end{tabular}
\end{table}

\begin{table}[]
\caption{MAGIC spectral fit parameters. Normalization is obtained at E = 146 GeV in units of 10$^{-10}$ TeV$^{-1}$ cm$^{-2}$ s$^{-1}$.}
\label{tab_spec}
\centering
\begin{tabular}{llll}
\hline
Data set & MJD 58029-30 & MJD 58057 & Low state \\
Eff. time [h]& 2.0 & 2.7 & 35.0 \\
Significance [$\sigma$]& 5.7 & 7.7 & 5.6 \\
Normalization  & 2.91 $\pm$ 0.62 & 3.22 $\pm$ 0.59 & 0.54 $\pm$ 0.13 \\
Spectral index & -3.86 $\pm$ 0.32 & -4.00 $\pm$ 0.27 & -3.52 $\pm$ 0.39 \\
\hline
 \end{tabular}
\end{table}

\begin{table}[]
\caption{Parameters for the jet-sheath model for $E_{\rm p,max}$=$10^{16}$.}
\label{tab_param}
\centering
\begin{tabular}{lcc}
\hline
State & MJD 58029-30 & Lower VHE \\
\hline
$B$ [G] & 2.6 &  2.6  \\
$E_{\rm min}$ [eV]& $3.2\times10^8$ & $2.0\times10^8$ \\
$E_{\rm br}$ [eV]& $7.0\times10^8$ & $9.0\times10^8$ \\
$E_{\rm max}$ [eV]& $8\times10^{11}$ & $8\times10^{11}$\\
$n_1$ & 2 & 2 \\
$n_2$ & 3.9 & 4.4 \\
$U_e$ [erg cm$^{-3}$] & $4.4\times10^{-4}$ & $3.6\times10^{-4}$ \\
$U_B$ [erg cm$^{-3}$]& 0.27 & 0.27 \\
$U_p$ [erg cm$^{-3}$] & $1.8$ & $0.7$\\
$P_e$ [erg s$^{-1}$] & $2\times10^{42}$ & $ 1.6\times10^{42}$ \\
$P_p$ [erg s$^{-1}$] & $8\times10^{45}$ & $ 3\times10^{45}$ \\
$P_B$ [erg s$^{-1}$] & $1.2\times10^{45}$ & $1.2\times10^{45}$ \\

\hline
 \end{tabular}
\end{table}

\section*{Acknowledgments}
We would like to thank the Instituto de Astrof\'{\i}sica de Canarias for the excellent working conditions at the Observatorio del Roque de los Muchachos in La Palma. The financial support of the German BMBF and MPG, the Italian INFN and INAF, the Swiss National Fund SNF, the ERDF under the Spanish MINECO (FPA2015-69818-P, FPA2012-36668, FPA2015-68378-P, FPA2015-69210-C6-2-R, FPA2015-69210-C6-4-R, FPA2015-69210-C6-6-R, AYA2015-71042-P, AYA2016-76012-C3-1-P, ESP2015-71662-C2-2-P, CSD2009-00064), and the Japanese JSPS and MEXT is gratefully acknowledged. This work was also supported by the Spanish Centro de Excelencia ``Severo Ochoa'' SEV-2012-0234 and SEV-2015-0548, and Unidad de Excelencia ``Mar\'{\i}a de Maeztu'' MDM-2014-0369, by the Croatian Science Foundation (HrZZ) Project IP-2016-06-9782 and the University of Rijeka Project 13.12.1.3.02, by the DFG Collaborative Research Centers SFB823/C4 and SFB876/C3, the Polish National Research Centre grant UMO-2016/22/M/ST9/00382 and by the Brazilian MCTIC, CNPq and FAPERJ. 
Part of this work is based on archival data, software or on-line services provided by the Space Science Data Center - ASI.

\end{document}